\documentclass{article}

\usepackage{PRIMEarxiv}

\usepackage[utf8]{inputenc} 
\usepackage[T1]{fontenc}    
\usepackage{hyperref}       
\usepackage{url}            
\usepackage{booktabs}       
\usepackage{amsfonts}       
\usepackage{nicefrac}       
\usepackage{microtype}      
\usepackage{lipsum}
\usepackage{fancyhdr}       
\usepackage{graphicx}       
\graphicspath{{media/}}     

\usepackage{multirow}%
\usepackage{amsmath,amssymb,amsfonts}%
\usepackage{amsthm}%
\usepackage{mathrsfs}%
\usepackage[title]{appendix}%
\usepackage{xcolor}%
\usepackage{textcomp}%
\usepackage{manyfoot}%
\usepackage{booktabs}%
\usepackage{algorithm}%
\usepackage{algorithmicx}%
\usepackage{algpseudocode}%
\usepackage{listings}%
\usepackage{tikz}

\usepackage[square,numbers]{natbib} 

\title{Solving the decision-making differential equations from eye fixation data in Unity software by using Hermite Long-Short-Term Memory neural network 

} 

\author{
  Kourosh Parand$^1$ \and Saeed Setayeshi$^2$ \and Mir Mohsen Pedram$^3$ \and Ali Yoonesi$^4$ \and Aida Pakniyat$^1$ \\
  *Department of Computer and Data Sciences, Faculty of Mathematical Sciences,\\
  Shahid Beheshti University, Tehran, Iran$^1$\\
  Department of Electrical and Computer Engineering, Amirkabir University, Tehran, Iran$^2$\\
  Department of Electrical and Computer Engineering, Kharazmi University, Tehran, Iran$^3$\\  
  Department of Neurosciences and Addiction Studies, Tehran University of Medical Sciences, Tehran, Iran$^4$\\
  \texttt{\{Corresponding author*: Kourosh Parand\}k\_Parand@sbu.ac.ir} \\
}

\begin{document}
\maketitle

\begin{abstract}

Cognitive decision-making processes are crucial aspects of human behavior, influencing various personal and professional domains. This research delves into the application of differential equations in analyzing decision-making accuracy by leveraging eye-tracking data within a virtual industrial town setting. The study unveils a systematic approach to transforming raw data into a differential equation, essential for deciphering the relationship between eye movements during decision-making processes. 

Mathematical relationship extraction and variable-parameter definition pave the way for deriving a differential equation that encapsulates the growth of fixations on characters. The key factors in this equation encompass the fixation rate $(\lambda)$ and separation rate $(\mu)$, reflecting user interaction dynamics and their impact on decision-making complexities tied to user engagement with virtual characters.

For a comprehensive grasp of decision dynamics, solving this differential equation requires initial fixation counts, fixation rate, and separation rate. The formulation of differential equations incorporates various considerations such as engagement duration, character-player distance, relative speed, and character attributes, enabling the representation of fixation changes, speed dynamics, distance variations, and the effects of character attributes.

This comprehensive analysis not only enhances our comprehension of decision-making processes but also provides a foundational framework for predictive modeling and data-driven insights for future research and applications in cognitive science and virtual reality environments.
\end{abstract}


\section{Introduction}\label{sec1}
Decision-making is an integral part of our personal and professional lives. Individuals may face various options in the decision-making process. Decision-making is often not a significant challenge in natural and everyday affairs. However, decision-making can be difficult at times due to factors such as time constraints, ambiguity, inability to discern reality, or many goal options \cite{rezvan}. Another point is that we make numerous daily decisions, some of which we may not even be aware of. Many of our daily decisions occur unconsciously. Decision-making aims to find an optimal solution to a problem and achieve the best logical outcome. In practice, many obstacles can lead to indecisiveness and an inability to discern the situation accurately. Therefore, the existence of tools that can assess the efficiency of decision-making indicates their importance. One of these tools is eye tracking. In this study, we will analyze decision-making accuracy using eye tracking. It's no wonder that the topic of decision-making has held a special place in various fields such as psychology, economics, and management for a long time \cite{bib10,bib11}. Therefore, extensive research has been conducted in this area, mostly focusing on examining decision-making results. One of the areas that has received significant attention is the relationship between individuals' behavioral habits and their decision-making. Indeed, certain behavioral habits can influence an individual's decision-making and lead them to make incorrect choices in response to the intended stimuli. Even factors not necessarily associated with an individual's behavioral habits, such as fear, anxiety, stress, and anger, can hurt decision-making and prevent proper decision-making. Therefore, accurate decision-making analysis can help unravel many cognitive issues \cite{bib2}. However, what is particularly important is how one can identify what decision an individual has made or will make by examining their various eye movements, and whether the decision thus made was correct. This paper has designed a virtual environment as an industrial town to investigate the accuracy of an individual's decision-making. Participants are asked to engage in several rounds of gameplay, assuming the role of a character equipped with various weapons in the game environment. Their objective is to identify and eliminate individuals referred to as bad. This game has two groups of individuals: bad and good officers. It is crucial that bad and good officers are indistinguishable in appearance, and their behavior is the only criterion for differentiating them. The player or participant must be able to shoot at the bad individuals. Naturally, the prerequisite for this task is for the player to distinguish between the behaviors exhibited by bad and good before taking action to eliminate the bad. 
The concept of decision-making here refers to an approach that compels individuals to choose\cite{bib14}. Therefore, even in cases where individuals are not consciously thinking and deciding in the traditional sense, they are making decisions simply by choosing one thing from among several options without realizing it \cite{bib3}. The crucial point in this regard is how well an individual makes logical and experiential decisions in significant events that they encounter, or how accurate their decision-making is because when an individual is faced with significant events that also involve a high level of risk, they must be able to make the right decision. For example, when a person is driving, their decision in the face of an immediate event is crucial. An incorrect decision in such a case can lead to unfavorable outcomes for the individual or others. There are also very sensitive professions in which decision-making plays a fundamental role, and sometimes making a wrong decision may have irreparable consequences. The importance of this issue is particularly evident in sensitive professions such as piloting. Decision-making forms an important and undeniable part of life, and therefore, various methods have always been sought to teach the correct decision-making approach\cite{bib13}. The method used here involves tracking eye movements during decision-making and comparing them with the correctness or incorrectness of the decisions made. There are various strategies to understand individuals' decision-making processes, including techniques such as EEG, which examine brain signals. However, we used eye tracking for this purpose in our proposed method \cite{drparand202311}. Like other methods, eye tracking and studying eye movements can be very effective in understanding individuals' decision-making processes. Aside from the geometric and rotational movements of the eyes (up and down, left and right), changes in pupil size in response to visual stimuli can also be considered. The opening and closing of the eyelids may also be considered in analyzing decision-making. However, the emphasis in this study has solely been on rotational eye movements. 

 In this article, an attempt has been made to design a game using Unity software, which can be used to analyze decision-making using eye fixation. We have modeled the existing data extracted from the eye tracker as a differential equation and solved it using the Hermite neural network method.
\subsection{Eye Detection}
One important feature of the human face is the eyes \cite{bib26}. The movement of the eyes plays a significant role in expressing needs, satisfaction, emotions, and communication \cite{bib15}. The importance of eye movement in personal perception and attention in the visual world has led to the necessity of explicit knowledge about the eyes to extract important information from their features \cite{bib4}. Eye tracking is a tool that enables the examination of eye movements. It identifies the eyes' status and movement when scanning the surroundings or simultaneously focusing on a specific subject. In natural binocular vision, both eyes look at the same point, and the brain's visual area combines the two received images into a three-dimensional image \cite{bib7}. Researchers work on identifying the eyes and their impact on eye position in the image and gaze point determination \cite{bib27}. There are three aspects to identifying the eyes: 
1. Whether the eyes are present or not. 
2. The accuracy of interpreting the eye's state.
3. The ability to identify the eye frame by frame in video images. 
The status of the eyes is usually measured by the pupil or the cornea. It is crucial to note that there is a significant difference between "gaze tracking" and "eye identification." In eye identification, the focus is on the location and position of the eyes, while gaze tracking aims to determine the path of the gaze. When eye tracking is used, the eye position in the image is identified, and information about the eyes and the head position and gaze direction is estimated\cite{bib32}. This information itself is also important in cases related to gaze tracking. 

Eye tracking has various types, which are divided into three main categories: 
1. Measurement of the movement of an object (specifically with a lens) that takes place by attaching it to the eyes. 
2. Another type of it is optical, without directly involving the eyes.
3. Another method involves electrical agents by placing electrodes around the eyes.
In the invasive method, the connection with the eyes is established by placing a lens on the cornea \cite{bib18}. In this method, continuous eye movements are measured, recording very sensitive data about eye movements. The movement of the eyes in horizontal and vertical directions is recorded. The second method is optical, utilizing light, usually infrared. Infrared light is irradiated to the eyes and reflected. This method does not involve direct interaction with the eyes. It is desirable and cost-effective\cite{bib8}. The third method is based on measuring the difference in electrical potential. This method uses electrodes placed around the eyes. In this method, the eyes act as a source of uniform electric potential, such that even if the eyes are closed or in a dark environment, the eye status is still detectable. The electrodes are generally bipolar, with the positive pole towards the cornea and the negative pole towards the retina. The received electrical signals from the two pairs of electrodes located around the eyes are analyzed using the EOG method. If the eyes move from a central position, the status of the electrodes changes in the retina and cornea. These bipolar changes are followed, affecting the results of measuring electrical potential in EOG. 

The rotation of the eye is controlled by six muscles from the eye to the brain\cite{bib17}. The muscles of the eyeball cause the movement of the eyeball. The eyeball can move along two horizontal axes without any displacement\cite{bib29}. There are three types of eye movements\cite{bib28}: 1) fixation, 2) saccade, and 3) pursuit. 
Fixation means the ability to keep the eyes uniformly focused without deviating from the target. Fixation allows maintaining a steady gaze without involuntary eye movements separating them from the target. In the fixed state, the eyes remain in a specific position for a certain period. In other words, fixation means stitching the eyes onto objects. Typically, when reading a text or scanning a scene, the human eye makes consecutive fixations for about 0.3 seconds each. Saccade refers to the sudden jump of the eye: rapid changes that occur voluntarily and intermingle with fixations, during which the eye moves from one subject to another. A precise saccade is defined based on the ability to see the surroundings of the main subject simultaneously, which may contain the next target. Saccades involve important visual mechanisms, including central vision focused on a target and peripheral vision that identifies the surroundings, allowing the brain to create a mental image based on the visual field. Saccades occur approximately three times per second between fixed intervals and last for about 0.3 seconds each. Pursuit refers to smooth and delicate tracking movements that maintain the image of a moving object on the retina. It involves subtle movements to keep the image on the fovea. Pursuit is the act of following objects with the eye and maintaining visual contact. With pursuit movements, the eye can track a target with a single smooth motion. This type of eye movement is essential for analyzing the speed and status of a moving object, and it holds particular importance in sports and driving. Researchers have extracted the features required in their experiments by utilizing the eye movements mentioned above. In some cases, two types of eye movements are used to express the features \cite{bib20}. Algorithms related to eye movements have been developed to extract data from regular images and videos, such as the AVT algorithm. This algorithm counts the number of pixels present in a central square region placed over the pupil of the eye to extract the fixation and saccade movements. Additionally, using statistical methods, a new method called Imap has been introduced to obtain fixation and saccade eye movements.

Eye tracker uses infrared sensors to identify eye characteristics and determine where a person is looking. Shortwave infrared technology is employed in i-tracing, utilizing advanced mathematical models to pinpoint the gaze. The technology aims for all stages to be automated while operating precisely and reliably in a wide range of environments\cite{bib19}. I-tracing has been utilized in numerous practical applications, broadly categorized as active and passive \cite{bib7}. 
The active method has been utilized in this study. In an active application, eye movement is used as a specific input for controlling a device. One of the areas where this is used is in gaming. In an active application, the user's gaze point is combined with other inputs such as keyboard and mouse, creating a new interactive mode. However, a passive application can analyze user attention without the need for invasive tools to analyze and dissect stimuli. In a passive application, only the data collected from the user's gaze is utilized. The collected data can be valuable for analyzing many stimuli and improving usage. In an active application, the focus is on controlling the device and using applications for gaming purposes. On the other hand, passive applications encompass cases that involve analysis for design or advertising purposes. However, the analysis goal in a passive application can also be a medical diagnosis, vehicle safety, or academic research. 

The Eye Tribe technology can accurately detect eye movement and pupil size within millimeter precision. Its average accuracy is approximately 0.5 degrees within the field of view. This means that The Eye Tribe can determine the position of the eye on the screen within a size of less than 10 millimeters. The technology calculates the gaze point and location where the individual is looking. This system is capable of integrating with other devices such as keyboards, mice, and touchscreens to create active applications. The data obtained from The Eye Tribe can be utilized in other technologies to enhance their functionality \cite{bib30}. By determining where the individual is looking, the Eye Tribe system extracts the location of the gaze point using information obtained from the user's face and eyes. The eye gaze coordinates are obtained in the form of $(x, y)$ coordinates, considering the screen that the user is looking at. To track the user's eye movements and extract gaze point coordinates, The Eye Tribe needs to be placed on the fine screen that the user is looking at. The properties related to The Eye Tribe are included in Table \ref{1}.  \vfill

\begin{table}[h]
\caption{The properties of the Eye tribe\cite{bib30}}\label{1}

\begin{centering}
\begin{tabular}{|c|c|}
\hline 
Sampling rate & 30Hz and 60Hz mode\tabularnewline
\hline 
Accuracy & 0.5° -- 1°\tabularnewline
\hline 
Spatial Resolution & 0.1° (RMS)\tabularnewline
\hline 
Latency & \textless 20ms at 60Hz\tabularnewline
\hline 
Calibration & 9, 12 or 16 points\tabularnewline
\hline 
Operating range & 45cm -- 75cm\tabularnewline
\hline 
Tracking area & 40cm x 30cm at 65cm distance (30Hz)\tabularnewline
\hline 
Screen sizes & Up to 24\textquotedblright{}\tabularnewline
\hline 
API/SDK & C++, C\# and Java included\tabularnewline
\hline 
Data output & Binocular gaze data\tabularnewline
\hline 
Dimensions (W/H/D) & 20 x 1.9 x 1.9 cm (7.9 x 0.75 x 0.75 inches)\tabularnewline
\hline 
Weight & 70g\tabularnewline
\hline 
Connection & USB3.0 Superspeed\tabularnewline
\hline 
\end{tabular}
\par\end{centering}
\end{table}
\vfill

In this paper, the sampling rate used was 30 Hz, and a 9-point scale was used for calibration. The maximum monitor size should be 24 inches, and the eye tracker should not be placed above or to the side of the monitor. The best placement is directly beneath the monitor. It is important to note that the eye tracker must be in a fixed position. The intended individual should be positioned in front of the monitor at an appropriate distance of 45-75 centimeters from the tracker and in the center. 
The information obtained from the eye tracker is stored in two ways: 1) Using an API and 2) Storing the data within the Unity application. The API is a tool that records the data obtained from the eye tracker through the server. The output obtained through this method is in the form of gaze coordinates, and this action takes place after the grading. In this method, it is possible to synchronize this console using settings within the eye tracker and Unity. The information obtained from this console includes the coordinates of the left and right eyes, the pupil center, and the pupil size. In the method used in this thesis, writing a C\# code within the Unity application and calling it directly is employed.

After these settings, the eye tracker is ready for use. To directly record the data available in the application while using Unity, a script called the "$FirstPersonController$" is defined, which is linked to the camera. In this script, the eye tracker information is recorded, providing the user with the eye coordinates, pupil center, and size. The data obtained from the eye is in screen coordinates and is defined in pixels.
 \section{Data extraction}
In this article, we performed the following steps to extract the data and convert them into a differential equation
\begin{itemize}
    \item The game was designed by Unity software
    \item Using the eye tracker for recording data 
    \item We convert the obtained data into a differential equation
    \item Finally, we solved this differential equation
\end{itemize}
\subsection{Game design}
There are various software available to design virtual environments as games. One of these software is Unity \cite{bib22}.

Unity is one of the newest game development software. Before version 2.5, this software was only available for Macintosh applications. Some of the features of this powerful engine are mentioned below:

\begin{itemize}
    \item A powerful, flexible editor that can execute user commands visually.
    \item Tools necessary for collaborative work, the ability to directly import files from software such as XSI, Maya, and Max, and update assets while running the software and creating games.
    \item Use the latest graphic technologies, high rendering speed, and all DirectX and OpenGL capabilities.
    \item Ability to create executable Exe outputs.
    \item One of the most important features of Unity is the ability to use AI and Unity simultaneously. Due to the high capabilities of Unity, AI can be recognized in the Unity environment.
\end{itemize}
Although Unity has only been around for about four years, it has provided users with a straightforward, beautiful, and efficient environment in this short period.
Unity is a popular and powerful game development platform that allows developers to create 2D, 3D, AR, and VR games and applications. It provides a user-friendly interface, and a wide range of tools and assets, and supports multiple platforms, including mobile devices, consoles, and PCs. Unity is widely used in the gaming industry and has a large community of developers and resources for learning and support \cite{bib31}. 
The Unity software is used for building video games for personal computers, game consoles, and mobile devices. This software was introduced at the Apple Worldwide Developers Conference in 2005 for the OSX operating system and has been under development since then. This game engine utilizes the MonoDevelop software, an open-source programming language editor. Unity is one of the game engines that supports popular programming languages, including C\# and JavaScript. In this study, an industrial environment was created using this software, where both good and bad characters, who are visually similar, are designed. 

\begin{itemize}
    \item One of the best features of the Unity engine that has made game development simple and enjoyable for game developers is the existence of a game object called Prefab. The game designed in this paper is an industrial town, using human models with a specific white skin texture and unique characteristics. These human objects are intended to be placed at 20 different points. When we want to change the characteristics of these individuals, making changes to each of them individually will be difficult. Therefore, using Prefab makes it easy to make changes. From now on, individuals are transformed into a Prefab, and making changes in the game will be very simple. 
    \item The most important part of the game is the game object, which can perform various tasks and includes components called Components. Each of these components is capable of performing a specific task. One of these is the Collider. The Collider is a part of the object that gives it a physical nature. For human objects, the best Collider is the Sphere Collider. This allows objects to detect collisions and have natural movements.
    \item Animator is a great feature of Unity. It gives motion to characters, and using various animations, different operations can be assigned to the character. 
    \item  Tags or labels are assigned to objects and Unity characters and are used when called upon.

\end{itemize}
The version of Unity used in this study is 5.1.1. The game environment is designed as an industrial town. The implementation of the industrial town is described below. 
\subsubsection{Industrial Town}
\begin{figure}[h]
\centering
\includegraphics[width=0.9\textwidth]{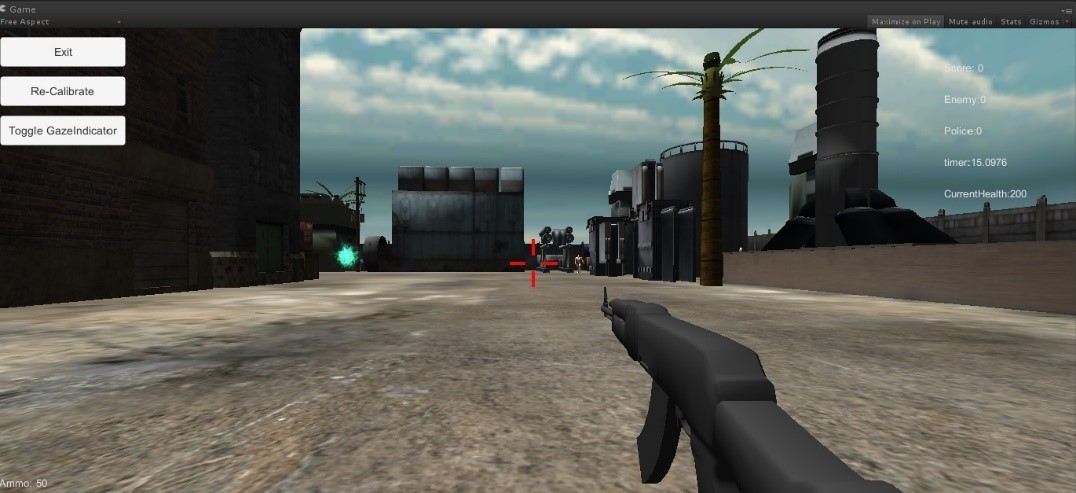}
\caption{The Industrial Town}\label{Industrial}       

\end{figure}

The space designed in Unity resembles an abandoned industrial town with buildings, towers, and derelict cars. Initially, it appears to be an abandoned town with no one present. However, after a few moments, the participant realizes that there are others in this town. The environment is somewhat designed to create fear and anxiety in the participant. The participant in this game is known as the first person or FPS, given a Glock pistol, and is required to complete the game within a designated time of around 5 minutes. Figure \ref{Industrial} shows the industrial town in this game. 

\begin{figure}[h]
\begin{centering}
\includegraphics[width=0.9\textwidth]{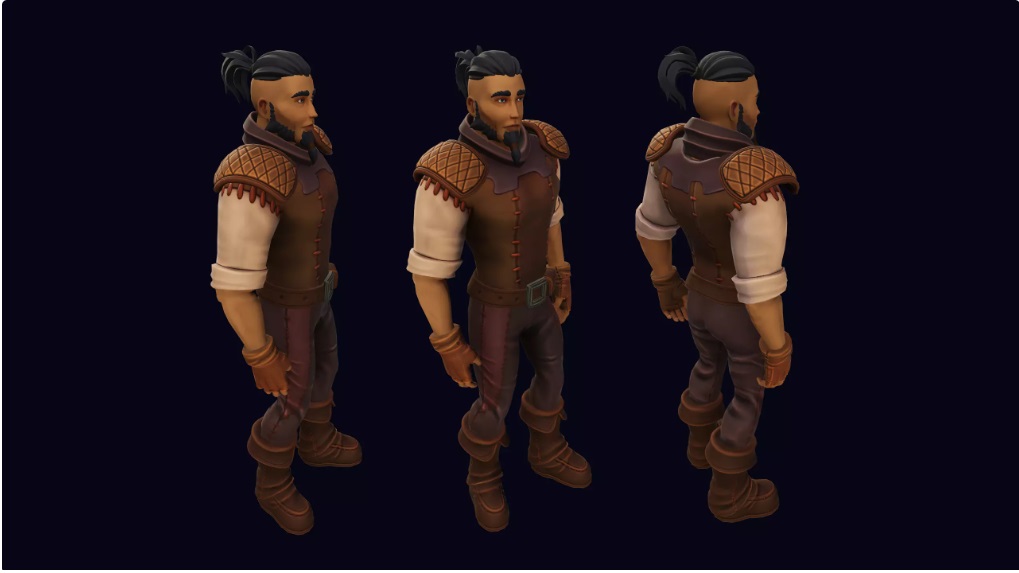}
\par\end{centering}
\caption{An example of a character used in Unity\cite{bib31}}\label{character}
\end{figure}

In the town, besides the participant, there are other individuals representing two groups. These two groups are named the good and the bad. In this game, the labels used for them are good and bad, respectively. The individuals representing the good and bad groups move within the town without causing harm to each other. The problem arises when they observe a new person who is unfamiliar to them. If the new person gets too close to them, these two groups change their current direction and move towards the new person. The behavior displayed by the two representatives when they see the new person is different from each other. Since these two groups have a very similar appearance, it will be very difficult to determine which one might be the bad and which one might be the police. Despite their strong resemblance, their behaviors towards the newcomer are different. The behavior of the friendly representatives is amicable, and they do not engage in any actions toward the newcomer other than meeting and leaving him. Meanwhile, the behavior of the hostile representatives is aggressive, and they start running toward the newcomer from a certain distance and then attack him, causing harm to the newcomer in each attack. Figure  \ref{character} shows an example of the character \cite{bib31}. 

Although the representatives of the two good and bad groups are physically larger than the newcomer, they are killed after two shots, and this applies equally to both groups. The newcomer, who is the participant or FPS, is asked to take the initiative to kill the bad representatives after identifying them. Shiny stones are placed in the environment for the participant. The significance of these stones is that they indicate the path to the participant, especially in areas where the likelihood of bad representatives being there is higher.

The ability given to the FPS in this game is to fight and run, which is interesting for many participants. The game story is also full of fresh ideas designed by experts in the fields of artificial intelligence and neuroscience. The occurrence of the game story is in an industrial town. The game story begins when all individuals from the two good and bad groups are at ease. The participant or FPS is the main character of the game, a character determined by the user. It becomes interesting when the participant sees the individuals in the game. The focus of the story is on the decisions made by the FPS. The other characters in the game, representatives of the two good and bad groups, display unpredictable behaviors toward the participant. One important consideration in the game is to prevent it from becoming tiring. Therefore, shiny stones are placed in the environment to prevent the participant from solely focusing on the end of their task and to draw them away from the environment, as well as to indicate the locations where bad may be present.

The main aim of this game is to kill the bad characters. As the FPS approaches the end of the game, the bad characters hide in specific locations. Due to the artificial intelligence of the characters, which fluctuates greatly, it is not certain how to move among them, as the bad characters may be behind the FPS without them noticing, or they may see the player from a distance and launch a collective attack on them. However, since decision-making is based on visual cues, the participant is required to eliminate bad with precision and complete confidence. This means that efforts have been made to prevent collective bad attacks, and this is done by scattering them. The weapon embedded for the bad is only hand-to-hand zombie attacks, which deplete the FPS's life with each strike.
Unity is a game engine that supports programming languages such as C\# and Java. As previously mentioned, Unity has utilized MonoDevelop, an open-source text editor software. In this game, the bad and good characters possess artificial intelligence. To make use of the AI capability for interacting with the prefabs in the game, the codes have been written in Unity's C\# script, and the Animator property has been used to animate the prefabs. What may be important for two categories of prefabs (good and bad) is that specific behaviors have been defined for both categories using a programming language. The behavior of both categories of prefabs before encountering the ordinary participant is based on the objectives defined for them, and they move within the environment. The behavior of these two groups becomes different when they see the participant at a defined distance. In this case, both prefabs change direction towards the FPS. The behavior of the good prefab, after seeing the FPS at a distance of 10 meters, is to move towards it and then leave the location. However, the behavior of the bad prefab is different: after reaching a distance of 10 meters from the participant, it rushes towards the participant, and then attacks at a distance of 5 meters. The programs written for player, bad, and good characters have all been implemented in Unity's C\# environment, and to provide artificial intelligence capabilities to the bad and good, in addition to the written codes, Animator has also been used. 
\subsubsection{Data collection from Unity}
In this study, two sets of data are collected: 1) Data obtained from Eye Tracker and 2) Data obtained from Unity. This section focuses on the collection of data obtained from Unity. To extract features, it is necessary to collect data from the environment and the eye. The data obtained from the Unity environment is as follows:
1. Whether it is a bad or a good character. 
2. The identifier of the bad and good. 
3. The coordinates where it was shot. 
4. The distance of the bad or good from the player at the time of shooting.
5. The coordinates where the bad or good was shot. 
6. The coordinates of the player when shooting the bad or good. 
7. The time when the bad and good were shot. The above items are recorded during game execution. These data indicate the participant's behavior in decision-making during the experiment. As mentioned before, the main goal is to determine whether the target character is good or bad, and then decide whether to eliminate them or not. It should be noted that these items are recorded after the shooting is done. 
\subsubsection{The Eye Tracker}

 As previously mentioned, in this paper, the need for two components to collect data is emphasized: 1) the virtual environment, and 2) eye-tracking. With the advancement of technology, a platform for research in cognitive sciences has been provided \cite{bib30}. Considering the diversity of cognitive science domains, the research methods have also become diverse and extensive. Experimental and empirical methods are widely used in this field, but alongside them, other research methods significantly contribute to cognitive studies. Methods such as computational and computer modeling can be mentioned, all of which share common characteristics: firstly, they are all reliant on empirical data, and secondly, an attempt is made to establish a common language with other domains, or at least a language that is translatable to other domains. This is achieved through experimentation where individuals' actions are recorded. Various techniques are used for this purpose, and one of these techniques is eye-tracking. The following section describes eye-tracking and its implementation in the virtual environment.

 The data obtained from the Unity environment that is useful for feature extraction includes 
\begin{enumerate}
    \item Eye coordinates $(x, y)$
    \item  Time
\end{enumerate}
The above aspects are recorded during gameplay in the Unity environment, showing users' eye movements during gameplay. While the data obtained from Unity is in three dimensions, the data obtained from the eye is in two dimensions. This is because the eye moves on the screen while individuals move in three-dimensional space during gameplay. Therefore, to simultaneously use eye and Unity data, Unity's spatial information is transformed into global coordinates.

One of the topics discussed in eye trackers is the analysis and interpretation of eye movements. There are different methods for analyzing eye movements. In some cases, only the feature of fixation or steadiness is considered, and in some research, both features of fixation and saccades are used. In this section, using the raw data obtained from the eye tracker and Unity, features of eye movements are extracted. Generally, we have two types of files: the file related to eye data and the file related to Unity data. In this thesis, the fixation feature is used for decision-making accuracy analysis. The work method for extracting features from raw data is as follows: The design and implementation of this method have been carried out in MATLAB 2011a software. The file related to the eye data is assumed to be in the time domain, and the file related to the Unity data is in the frequency domain. With this assumption, there are three general stages: 
\begin{itemize}
    \item First stage: The raw data obtained from eye movements is in the form of a matrix containing coordinates $(x, y)$ and time. The fixation identification algorithm operates based on two parameters: time and distance. Comparisons are made between two consecutive rows, and each row is compared to the next row in terms of (1) Euclidean distance and (2) threshold comparison. As mentioned earlier, the information obtained from the eye tracker is extracted from the screen and is in two dimensions. The type of information obtained from the eye is in pixels.  
The Euclidean distance $d$ between two consecutive $(x, y)$ rows, as obtained from \ref{eq1}, should be less than 5 pixels, and the time difference between two consecutive rows should also be less than a threshold, which is assumed to be 0.1 here. If the stated conditions are met, a fixation matrix, named "fix" in this program code, is constructed. The first stage is based solely on eye data.

\begin{equation}\label{eq1}
d=\sqrt{(y_{2}-y_{1})^{2}-(x_{2}-x_{1})^{2\\}}.
\end{equation}
The fixation matrix, or "fix," consists of 6 columns, including the fixation number, start and end times of the fixation, the duration of the fixation, the average coordinates of x and y on the screen, and the time interval between consecutive fixations. 
\item Second stage: By using the information obtained from the unity matrix, another matrix called "$unityfile$ " is formed. This matrix indicates in which fixation each shot occurred in the code. An important point to note is that the unity file itself is divided into three files. In the first file, the time of bullet impact on the character is recorded. In the second and third files, respectively, information about the bad and the good, as well as their distance from the player at each time, is recorded. The basic comparison used in this algorithm is the comparison of times between the files. In this stage, the start time and end time of the fixation are compared with the time of bullet impact on the character. If the time in this comparison matches, the "$unityfile$ " matrix, which consists of seven columns, is formed. The "$unityfile$ " matrix includes the following: the fixation number in which the shot occurred, the moment of the shot within which fixation, the start and end times of the fixation, the duration of the fixation, the character's goodness or badness, and the character's number. 
\item Third stage: Using the "$unityfile$ " matrix, it is determined how many fixations occurred on each character and how long each fixation lasted. The goal at this stage is to understand the behavior of characters that have been shot at concerning the player. As mentioned before, good and bad characters looked similar in appearance, and the only way to distinguish between good and bad characters was their behavior. The behavior of bad characters towards the player was not friendly. Both characters changed direction when they saw the player. However, the behavior of bad characters after changing direction was to run and attack the player. The moment of running is the moment when the player recognizes that the target is a bad character and needs to shoot. At this stage, the character's status at the time of the shot is examined as follows: Using "$unityfile$ ," another matrix is created in which it is specified whether the character shot. The fixation that occurred was good or bad, as well as its number. By using this matrix, the number of fixations at the time of the shot and the duration that the player looked at the target is obtained. By comparing other unity files and the obtained matrix, a new matrix is created. The resulting matrix is compared with the unity matrices that indicate the character's status, and ultimately a matrix called "learn" is created, consisting of five columns. The matrix from which the corresponding differential equation is extracted is the "learn" matrix. In some cases, the player may shoot without identifying the character. Since no fixation movement occurred in this action, it is considered an irrelevant option in feature extraction and hence is not recorded in the output. Therefore, this algorithm can also identify cases where the player has been targeted by chance without identifying the character's behavior.

\end{itemize}

\subsubsection{Test sample}
The necessary data for the experiment was collected from the participants who were present in the experiment. There were 17 participants, including 9 males and 8 females, aged between 20 and 40 years. The participants were asked to position themselves in a suitable location, 60 centimeters away from the computer and facing the eye tracker. After the classification was performed by the eye tracker, the gaming environment was explained to them, and they were asked to shoot based on the characters' behavior. The test took approximately 5 minutes for each participant. The eye tracker used was a high-quality commercial eye tracker, known as "$EyeTribe$," and the sampling rate used in this experiment was 30 Hz. The participants had a healthy vision and were all members of the National Library of Tehran. The stimulus used in this paper was a virtual environment designed in Unity software. 
Selecting appropriate features for analysis and examination is of utmost importance. As mentioned in the previous sections, the "learn" matrix is the final feature extraction matrix and includes the following: 
\begin{enumerate}
    \item The number of fixations on characters (good and bad),
    \item The duration of time the user fixated on the target,
    \item The distance of the character relative to the player,
    \item The speed of the character relative to the player.
\end{enumerate}
The basis of the feature extraction implemented in this algorithm is the examination of eye fixation movement. The goal of this algorithm is to identify the fixation movement in each player's shot towards the characters, taking into account the characters' behavior towards the player and eliminating chance factors. 

\section{Differential equation}

 The general process to transform data into a differential equation is as follows: 
 \begin{enumerate}
     \item Extraction of the mathematical relationship between variables: Firstly, we need to examine the data and find the mathematical relationship between the variables. For this, we can use various methods such as regression analysis, content analysis of the data, or other statistical methods.
     \item  Definition of variables and parameters: After examining the data and extracting the mathematical relationship, we must consider the next step. This step involves defining the variables and parameters that play a role in the differential equation.
     \item Extraction of the differential equation: Using the mathematical relationship and the definition of variables and parameters, we can construct the desired differential equation. At this stage, it is crucial to pay attention to the types of variables (such as dependent and independent variables) and their influence on each other.
     \item  Parameter adjustment and equation evaluation: After defining the differential equation, we need to adjust the parameters of the equation and make sure that the equation can accurately describe the data. For this, we can use analytical or numerical methods to match the data with the model. Following these steps will help us construct a differential model to describe the mathematical relationship in the data and use it for prediction or data analysis.

 \end{enumerate}

According to the extracted data from the eye tracker, we can derive its differential equation\cite{bib18}. The differential equation that has been considered is the equation for the growth of the number of fixations on a character. This equation is based on two main factors: the fixation rate and the separation rate.

We assume, $\lambda$ represents the fixation rate (the number of new fixations per unit of time) and $\mu$ represents the separation rate (the number of fixations that occurred but the user was inactive per unit of time).

The separation rate $(\mu)$ represents the number of fixations that occur on the character per unit of time but the user temporarily gets tired of it and leaves. This rate indicates how much the user gets exhausted with the character and abandons it.

The differential equation models the change in the ratio of fixations $(dN(t)/dt)$ concerning time $(dt).$ This change in the ratio of fixations is equal to the difference between the fixation rate$ (\lambda) $and the separation rate $(\mu).$

In essence, the differential equation shows that the growth rate of fixations is equal to the difference between the fixation rate and the separation rate. If the fixation rate is greater than the separation rate, the number of fixations increases, and if the separation rate is greater than the fixation rate, the number of fixations decreases.

To solve the differential equation, we need initial conditions, i.e., the number of fixations at an initial time $(e.g., t=0) $ denoted by $ N0$. Additionally, for a complete solution of the equation, we need to know the fixation rate $(\lambda)$ and the separation rate $ (\mu)$.

With the provided initial conditions and information about the fixation and separation rates, we can solve the differential equation and continue my previous explanations:

Based on the information we have provided, we can express the differential equation in a general form. If we denote the number of fixations as $N(t)$, the duration during which the user is engaged. Fixation occurs as T, the distance from the character to the player as D(t), the character's speed relative to the player as $V(t)$, and a variable called $G(t)$ to indicate whether the character is good or bad $G(t) = 1$ for being good and $G(t) = 0$  for being bad). 
In short, it can be stated as follows.:
\begin{itemize}
    \item $N(t)$: the amount of time the user gazed at the target and fixation occurred
    \item $D(t)$: distance of the character to the player
    \item $V(t)$: Speed of the character relative to the player
    \item $G(t)$: Goodness of the desired character (1 for bad and 0 for good)
\end{itemize}

Given these explanations, we can express the differential equations as follows:

\begin{equation}\label{eq3}
\frac{dN(t)}{dt}=\lambda G(t) - \mu N(t),
\end{equation}
\begin{equation}\label{eq4}
\frac{dV(t)}{dt}=V(t)\frac{dV(t)}{dt} - kV(t),
\end{equation}
\begin{equation}\label{eq5}
\frac{dD(t)}{dt}=V(t) - N(t),
\end{equation}
\begin{equation}\label{eq6}
\frac{dG(t)}{dt}=G(t)\frac{dG(t)}{dt} - mG(t),
\end{equation}

Here, $\lambda$ represents the fixation rate, which signifies the number of new fixations per unit of time. Additionally, $\mu$ represents the separation rate, which represents the number of fixations that occur but the user gets tired of and does not return to the game.

The first differential equation \label{eq3} models the change in the ratio of fixations $(dN(t)/dt) $ concerning time $(dt)$.  This equation describes the rate of change of the number of fixations $(N(t))$ concerning time $(t)$. The term on the left-hand side, $(dN(t)/dt) $ represents the derivative of $N(t)$ for $t$, which quantifies how the number of fixations changes over time.

The right-hand side of the equation consists of two terms. The $\lambda$ term represents the fixation rate, which denotes the rate at which new fixations occur. It reflects how attractive or engaging the character is to the users. The $\mu N(t)$ term represents the separation rate, where $\mu$ is the separation rate itself, and $N(t)$ is the current number of fixations. This term accounts for the fixations that cease to occur due to users becoming temporarily tired, losing interest, or getting distracted from the character.

The equation essentially states that the growth rate of fixations is equal to the fixation rate minus the separation rate times the current number of fixations. If the fixation rate is greater than the separation rate times the number of fixations, the number of fixations will increase over time. On the other hand, if the separation rate is greater, the number of fixations will decrease.

To solve this differential equation, we must specify initial conditions, such as the number of fixations at a particular initial time $(N(0))$. Additionally, we would need to know the values of the fixation rate $\lambda$ and the separation rate $\mu$. With these parameters, we can use various techniques such as separation of variables, integrating factors, or numerical methods to solve the equation and obtain a solution that describes the growth of fixations on the character over time.
This equation \ref{eq4} represents a self-reinforcing or positive feedback process. Positive feedback occurs when a change in a variable amplifies itself, leading to further changes in the same direction. In this case, the rate of change of $V(t) (dV(t)/dt)$ is multiplied by $V(t)$ itself, which means that the rate of change is proportional to the current value of $V(t)$. This self-reinforcing behavior can result in rapid growth or decay of the variable, depending on the initial conditions and the value of the constant $k$.
The second term,$ -kV(t)$, acts as a damping factor or a damping term in the equation. It introduces a negative influence that opposes the self-reinforcing effect. The term $-kV(t)$ represents a damping force that is proportional to the current value of $V(t)$ but acts in the opposite direction. This damping force can slow down the growth or decay of $V(t) $and ultimately bring it towards a steady state or equilibrium.

The value of the constant $k $ determines the strength of the damping effect. A larger value of $ k$ implies a stronger damping force, leading to a faster convergence to equilibrium. On the other hand, a smaller value of $k $ determines self-reinforcing behavior to dominate, resulting in more pronounced growth or decay.

Furthermore, the second differential equation represents the change in the distance from the character to the player $(dD(t)/dt) $concerning time $(dt)$. $G(t)$ represents a function in equation \ref{eq6} that changes concerning time $t$. The equation describes how the rate of change of $G(t)$, denoted as $dG(t)/dt$, is influenced by two factors: the input function $h(t)=G(t) (dG(t)/dt) $ and the decay term $mG(t)$.
The term $h(t)$ represents an external input or driving force that affects the rate of change of $G(t)$. It could be a function of time or any other relevant variables in the problem. The specific form of $h(t)$ depends on the nature of the problem and can be determined based on the input requirements or conditions.
The second term, $mG(t)$, represents the decay of $G(t)$ over time. The parameter $m$ determines the decay rate or how fast $G(t)$ decreases. A larger value of $m$ indicates a faster decay, while a smaller value of $m$ leads to a slower decay. The decay term $mG(t)$ represents the loss or decrease of the quantity $G(t)$ over time.
To solve the differential equation and determine the behavior of $G(t)$, we typically need to provide an initial condition. This condition specifies the value of  $G(t)$ at a specific time, often denoted as $G(0)$. By integrating the differential equation with the given initial condition and solving for $G(t)$, we can obtain the function that represents the variation of $G(t)$ over time.
In summary, $G(t)$ in the given equation represents a function that changes over time, and its rate of change is influenced by an external input $h(t)$ and a decay term $mG(t)$. By solving the differential equation with appropriate techniques and considering the initial condition, we can determine the behavior of $G(t)$ over time.

Based on the provided data, the initial conditions for the differential equations can be stated as follows:

\begin{itemize}
    \item Fixation Count ($N(t)$) at the initial time $t=0$: $N(0) = 70$
    \item Character Speed ($V(t)$) relative to the player at the initial time $t=0$: $V(0) = 48$
    \item Distance ($D(t)$) between the character and the player at the initial time $t=0$: $D(0) = 47$
    \item Character Goodness ($G(t)$) at the initial time $t=0$:  $G(0) = 2$
\end{itemize}

Based on the provided data, we can calculate the values of $\mu$, $\lambda$, $m$, and $k$ as follows:

\begin{enumerate}
    \item $\mu$: The separation rate is the average rate at which fixations cease to occur due to users becoming temporarily tired, losing interest, or getting distracted from the character. We can calculate this by dividing the number of fixations that end by the total fixation period:

$\mu= \frac{Number of fixations that end}{Total fixation period}$

The given data shows that fixations end when the character is classified as "bad" (1). Based on this, we can calculate the total fixation period as the sum of the "period of fixation" for each fixation that ends. For example: counting only the fixations that end and summing their respective periods, we get:

Total fixation period = $14.7407 + 1.3207 + 1.3207 + 0.367 + 0.1745 = 18.9239$

The number of fixations that end is equal to the number of fixations classified as "bad" (1):

Number of fixations that end = 6

Therefore, we can calculate the separation rate $\mu$ as:

$\mu = \frac{6}{18.9239} $

    \item $\lambda$: The fixation rate represents the average rate of new fixations. We can calculate this by dividing the total number of fixations by the total fixation period:

$\lambda = \frac{Total number of fixations}{Total fixation period}$

In the given data, we can calculate the total number of fixations by counting the number of entries in the data set:

Total number of fixations = 36

Using the total fixation period calculated in the previous step, we can calculate the fixation rate $\lambda$ as:

$\lambda = \frac{36}{18.9239} $
    \item $m$: The decay rate represents how fast the goodness of the desired character decays over time. Based on the given data, we can observe that the goodness of the character changes from "good" (0) to "bad" (1) during some fixations. We can count the number of fixations where the character changes from "good" to "bad" and divide it by the total number of fixations:

$m = \frac{Number of fixations where the character changes from "good" to "bad"}{Total number of fixations}$

We can count 5 fixations in the given data where the character changes from "good" to "bad". Therefore, we can calculate \$m\$ as:

$m = \frac{5}{36} $
    \item $k$: The constant $k$ determines the strength of the damping effect in the equation describing the character's speed. We can express the formula for calculating k as follows:

$k=(\frac{\min distancevalue}{\min distanceinthedata})+(\frac{\max speedvalue}{\max speedinthedata})$

By substituting the calculated values into the variables, the formula becomes:
       $k = (3.910988834 / 3.910988834) + (78.16336 / 78.16336) $

\end{enumerate}
To summarize, based on the provided data, we can calculate the values of $\mu$, $\lambda$, $k$ and $m$ as approximately:

\begin{itemize}
    \item $\mu = 0.3170$
    \item $\lambda = 1.9036$
    \item $m =0.1389$
    \item $k=2$
\end{itemize}

Using these differential equations, we can solve them numerically or analytically to obtain more accurate results regarding the behavior of fixations, the distance from the character, and the character's speed relative to the player.
By solving the equation, we can gain insights into how the number of fixations changes over time and understand the impact of the fixation rate and separation rate on the growth or decline of fixations. This information can be valuable for analyzing user behavior, optimizing character design, or predicting the character's popularity. 
There are various numerical methods for solving differential equations, which are chosen based on the properties and requirements of the problem. Some of these methods include:

\begin{itemize}
    \item \textbf{Finite Difference Method: }In this method, partial or overall derivatives of the differential equation are approximated differential using approximate formulas. Then these differential equations are solved with the existing boundary conditions. This method is suitable for differential problems with specific boundary conditions \cite{bib33}.
    \item \textbf{Finite Element Method: }In this method, the space is divided into smaller elements, and the differential equations in each element are solved as algebraic equations. By combining the solutions of the elements, the final solution is obtained. This method is suitable for problems with variable domains and diverse boundary conditions \cite{bib34,shen}.
    \item \textbf{Spectral Methods:} In this method, approximation functions created by spectral functions such as orthogonal\cite{drparand2004} are used to approximate the exact solutions of the differential equation. This method is suitable for problems with continuous exact function solutions\cite{khater2021,drParand2018-2}
\end{itemize}
In addition to different numerical methods, machine learning  \cite{drparand2023} can also be employed to solve differential equations, where a model is trained to learn a relationship between the inputs of the differential equation and its expected outputs (usually the values of the variables that the equation solves for)\cite{drparand20232}. Support Vector Machine $(SVM)$ methods can also be used to solve differential equations as a machine learning approach. In this approach, data consisting of inputs and outputs of the differential equation are used for the $SVM$ model to automatically learn the relationship between the inputs and outputs\cite{drparand2022}.  The $LS-SVR$ algorithm is a variation of Support Vector Machine $(SVM) $ models used for regression tasks.
In this method, the goal is to learn a non-linear function from the data that can provide the best fit to the data. To achieve this, $LS-SVR$ aims to minimize the distance between the training data points and the mean squared error of predictions\cite{drparand20224}
This algorithm utilizes two types of support variables: convex variables and penalty functions to minimize the estimation risk. This algorithm optimizes the problem of solving differential equations as an optimization problem.
This method seeks to minimize an objective function such as the squared error in the difference between actual and predicted values. This method also has a high capability of learning non-linear patterns and can be used to solve differential equations effectively \cite{drparand2022}.
The $LS-SVR$ is a powerful and efficient method that can be employed for solving differential equations and other prediction and regression tasks \cite{drparand202311}. 
 
The proposed model is typically trained using training data such as exact numerical solutions or empirical data (if available), and then used to predict input values (the differential equation) accordingly.
For instance, in the code I presented, a $LSTM$ model is used to predict the values of $N$, $V$, and $G$ with the help of a differential equation. The $ LSTM$ model can learn complex and time-sensitive patterns from the data provided and predicts the values of $N$, $V$, and $G$ over time based on the training data.

Given the complexity of differential equations and your data, a combined approach of approximate methods and machine learning may provide the best solution for the specific problem at hand.

 \section{Methodology}
 Solving differential equations using machine learning methods is considered a novel and powerful approach. In this method, artificial intelligence models such as neural networks, decision trees, support vector machines, and others are used to automatically and intelligently provide algorithms for solving differential equation problems. In a differential equation, there is a boundary problem with a specified start and initial values. Here, neural networks can be trained to analyze the data of the problem, and after training, they can accurately and quickly calculate the solution with high accuracy. The benefits of using machine learning methods for solving differential equations include high accuracy, a more comprehensive ability to solve complex problems, the ability to solve problems on a large scale, and process automation. However, the main challenge of this method is the need for a large amount of training data to train machine learning models \cite{drparand20233}.
 Recurrent Neural Networks $ (RNNs)$  \cite{bib9} are a type of artificial neural network architecture commonly used for sequential data processing tasks. In the context of solving differential equations, $RNNs$ can be applied to model dynamic systems and predict their behavior over time. Here is how $RNNs $ can be utilized for solving differential equations:

\begin{enumerate}
    \item Sequence Modeling: $RNNs$ are well-suited for capturing patterns in sequential data, making them useful for modeling the behavior of dynamic systems described by differential equations. By feeding sequential data into the $RNN$, it can learn the underlying patterns and relationships within the data, which can then be used to predict future states of the system.
    \item Time Series Prediction: For differential equations involving time-dependent variables, $RNNs$ can be trained on time series data to predict future states of the system. learning the temporal dependencies in the data, $ RNNs $ can effectively model the evolution of the system over time, making them useful for predicting the solutions of differential equations at different time steps.
    \item Long Short-Term Memory $(LSTM)$ Networks:$ LSTM $networks, a type of $RNN$ with memory cells designed to capture long-term dependencies, can be particularly useful for solving differential equations that involve complex dynamics and long-range dependencies. $LSTM$ networks excel at learning and retaining information over extended sequences, which can be beneficial for modeling the behavior of systems described by differential equations.
    \item Training and Optimization: To apply $ RNNs $ for solving differential equations, the network parameters need to be optimized to minimize the error between the predicted and actual solutions. This typically involves training the$ RNN$ on a dataset of known solutions to differential equations and adjusting the network weights through backpropagation to improve prediction accuracy.
\end{enumerate}
In summary, $RNNs,$ especially variants like $LSTM $ networks, can be effective tools for solving differential equations by leveraging their ability to model sequential data, capture temporal dependencies, and predict future system states.

We used the Long Short-Term Memory neural network with Hermite activation function to solve this coupled differential equation.  Using the code and the obtained predictions, we can provide an analysis of the future behavior of $N(t)$, $V(t)$, $D(t)$, and $G(t)$. By examining the predicted values' graphs and their temporal changes, we can identify clear patterns and trends. For example, by analyzing the graphs, we can determine whether $N(t)$ (the user's level of focus on the objective) is dependent on $G(t)$(the quality of the objective) or not. Does the value of $N(t)$ increase with an improvement in the objective's quality and decrease when the objective is poor?
Similarly, we can investigate the variations in $V(t)$ (speed) and determine whether these changes are influenced by speed itself or other variables. By observing the predicted values for $D(t)$ (character distance from the player), we can see if this value is influenced by speed or the user's level of focus on the objective. Lastly, we can examine the changes in $G(t)$ (objective quality) and gain insights into its behavior over time. By analyzing these patterns and trends, we can arrive at useful conclusions about the system's behavior and the interplay between variables.
\subsection{Hermite Functions}
In this section, we consider the properties of Hermite functions.
$\widetilde{H}_{n}(x)$ are the normalized Hermite functions of degree $n$, which describe the properties of Hermite functions \citep{shen,drParand2018-2}.
\begin{equation}
\widetilde{H}_{n}(x)=\frac{1}{\sqrt{2^{n}n!}}e^{\frac{-x^{2}}{2}}H_{n}(x),\quad n\geq0,\:x\in\mathbb{R}.
\end{equation}

The formula for orthogonal relation for Hermite functions is as follows:

\begin{equation}
\int_{-\infty}^{+\infty}\widetilde{H}_{n}(x)\widetilde{H}_{m}(x)=\sqrt{\pi}\delta_{mn},
\end{equation}

where $\delta_{mn}$ is the Kronecker delta function. Hermite functions have a recurrent relation defined in the  $(-\infty,+\infty)$ domain

\begin{gather}
\widetilde{H}_{n+1}(x)=x\sqrt{\frac{2}{n+1}}\widetilde{H}_{n}(x)-\sqrt{\frac{n}{n+1}}\widetilde{H}_{n-1}(x),\quad n\geq1,\nonumber \\
\widetilde{H}_{0}(x)=e^{\frac{-x^{2}}{2}},\;\widetilde{H}_{1}(x)=\sqrt{2}xe^{\frac{-x^{2}}{2}}.
\end{gather}

Use the Hermite functions' recurrence relation and formula to get the result

\begin{equation}
\widetilde{H^{\prime}}_{n}(x)=\sqrt{2n}\widetilde{H}_{n-1}(x)-x\widetilde{H}_{n}(x)=\sqrt{\frac{n}{2}}\widetilde{H}_{n-1}(x)-\sqrt{\frac{n+1}{2}}\widetilde{H}_{n+1}(x),
\end{equation}

and it becomes

\begin{equation}
\int_{-\infty}^{+\infty}\widetilde{H^{\prime}}_{n}(x)\widetilde{H^{\prime}}_{m}(x)dx=\begin{cases}
-\frac{\sqrt{n\pi(n-1)}}{2}, & m=n-2,\\
(n+\frac{1}{2})\sqrt{\pi}, & m=n,\\
-\frac{\sqrt{\pi(n+1)(n+2)}}{2}, & m=n+2,\\
0, & Otherwise.
\end{cases}
\end{equation}

\begin{equation}
\tilde{P}:\{u:u=e^{\frac{-x^{2}}{2}}\nu,\;\forall_{\nu}\epsilon P_{N}\},
\end{equation}
    
 Where $P_{N}$ represents the Hermite polynomials of degree $N$.   
    
 \subsection{Long-Short Term Memory}

$LSTM$ is a type of recurrent neural network $(RNN)$ architecture designed for the vanishing gradient problem in traditional $RNNs$. It is particularly effective in modeling and predicting sequential data because it captures long-term dependencies.

\begin{figure}[h]
\begin{centering}
\includegraphics[width=0.9\textwidth]{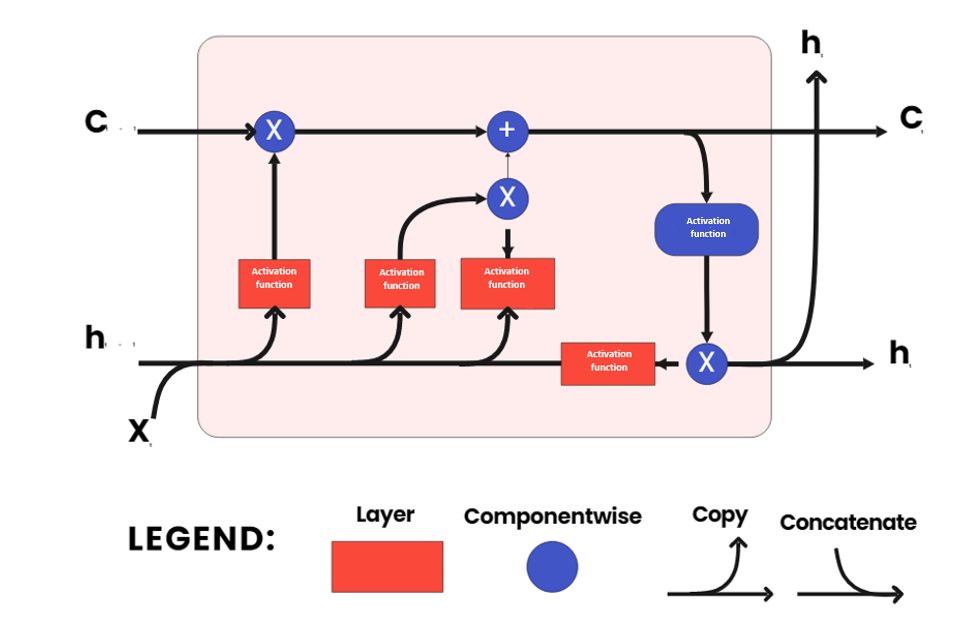}
\par\end{centering}
\caption{LSTM neural network}\label{LSTM}
\end{figure}

The $LSTM$ unit consists of several key components, including the input gate $
i_{t}$, forget gate $
f_{t}$, output gate $
o_{t}$, and cell state $
c_{t}$. These gates and the cell state are updated at each time step based on the input, previous hidden state, and previous cell state.  The $LSTM$ neural network  Components are shown in figure \ref{LSTM}.

The formulas for the $LSTM $ are as follows:

 Input Gate $(i_t)$: The input gate determines how much new information should be stored in the cell state $c_t$ based on the current input $x_t$ and the previous hidden state $h_{t-1}$.
 \begin{equation}
       i_t = activationfunction(W_{xi} \cdot x_t + W_{hi} \cdot h_{t-1} + W_{ci} \cdot c_{t-1} + b_i),
\end{equation}

Forget Gate $(f_t)$: The forget gate determines how much information from the previous cell state $c_{t-1}$ should be forgotten or discarded. It takes into account the current input $x_t$ and the previous hidden state $h_{t-1}$ and produces an activation function. 
\begin{equation}
 f_t = activationfunction(W_{xf} \cdot x_t + W_{hf} \cdot h_{t-1} + W_{cf} \cdot c_{t-1} + b_f),
\end{equation}
 Output Gate $(o_t)$:  The output gate determines how much of the cell state $c_t$ should be outputted as the hidden state $h_t$. It considers the current input $x_t$, the previous hidden state $h_{t-1}$, and the current cell state $c_t$ to generate values using the activation function. 
\begin{equation}
    o_t = activationfunction(W_{xo} \cdot x_t + W_{ho} \cdot h_{t-1} + W_{co} \cdot c_{t} + b_o),
\end{equation}

Cell State  $(c_t)$: The cell state is responsible for storing and propagating information throughout the LSTM unit over time. It is updated based on the input gate, forget gate, and the current input $x_t$. The input gate determines the amount of new information to be added to the cell state, and the forget gate determines the amount of previous information to be retained or discarded. The cell state is also modified by applying the activation function to the weighted sum of the current input, previous hidden state, and bias terms. 

\begin{equation}
    c_t = f_t \odot c_{t-1} + i_t \odot activationfunction(W_{xc} \cdot x_t + W_{hc} \cdot h_{t-1} + b_c),
\end{equation}

Hidden State $(h_t)$: The hidden state is the output of the $LSTM$ unit at each time step. It carries short-term dependencies and is influenced by the cell state. The hidden state is computed by applying the output gate to the activation function of the cell state. 
\begin{equation}
  h_t = o_t \odot activationfunction(c_t),
\end{equation}

where $x_t$ represents the input at time step $t, h_t$ is the hidden state at time step $t,W$ and $b$ are weight matrices and bias vectors, $\odot$ represents element-wise multiplication and Hermite functions are used as the activation function in this paper.

These formulas allow the $LSTM$ unit to control the flow of information through the input, forget, and output gates, and update the cell state accordingly. The cell state retains long-term information, while the hidden state captures the short-term dependencies and is passed to the next time step.

$LSTM$'s ability to selectively retain or discard information over long sequences makes it highly effective in various tasks, such as natural language processing, speech recognition, and time series prediction.

The $LSTM$ model is shown as follows \ref{fig:lstm_model}: 

\begin{figure}[H]
\centering
\begin{tikzpicture}[node distance=2cm, every node/.style={draw,align=center, text width=1cm}]

\node (input1) {Input Sequence};
\node (input2) [right of=input1] {Input Sequence};
\node (input3) [right of=input2] {Input Sequence};

\node (lstm1) [below of=input1] {LSTM Layer 1};
\node (lstm2) [below of=input2] {LSTM Layer 2};
\node (lstm3) [below of=input3] {LSTM Layer 3};

\node (output1) [below of=lstm1] {Output Layer};
\node (output2) [below of=lstm2] {Output Layer};
\node (output3) [below of=lstm3] {Output Layer};

\node (loss1) [below of=output1] {Loss Function};
\node (loss2) [below of=output2] {Loss Function};
\node (loss3) [below of=output3] {Loss Function};

\draw[->] (input1) -- (lstm1);
\draw[->] (input2) -- (lstm2);
\draw[->] (input3) -- (lstm3);

\draw[->] (lstm1) -- (output1);
\draw[->] (lstm2) -- (output2);
\draw[->] (lstm3) -- (output3);

\draw[->] (output1) -- (loss1);
\draw[->] (output2) -- (loss2);
\draw[->] (output3) -- (loss3);

\draw[dashed] ([xshift=-0.8cm]input1.north west) rectangle ([xshift=0.8cm]input3.south east);
\draw[dashed] ([xshift=-0.8cm]lstm1.north west) rectangle ([xshift=0.8cm]lstm3.south east);
\draw[dashed] ([xshift=-0.8cm]output1.north west) rectangle ([xshift=0.8cm]output3.south east);
\draw[dashed] ([xshift=-0.8cm]loss1.north west) rectangle ([xshift=0.8cm]loss3.south east);

\end{tikzpicture}
\caption{LSTM Model Architecture}
\label{fig:lstm_model}
\end{figure}
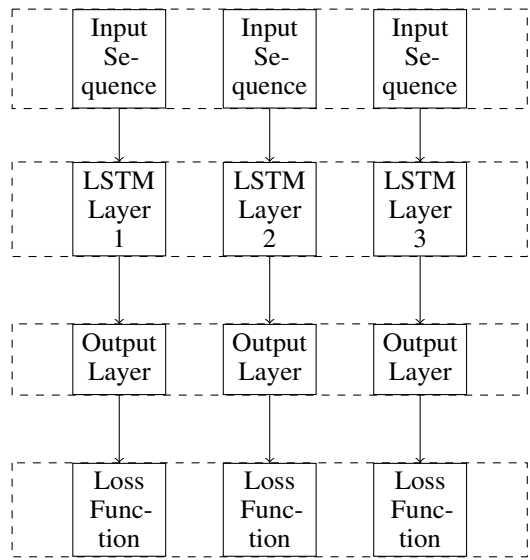

The $LSTM$ model processes sequences by iteratively applying the$ LSTM$ cells to each sequence element while maintaining an internal hidden state that carries information from previous steps. The hidden state serves as a memory that allows the model to capture long-term dependencies.

During training, the $LSTM$ model is optimized to minimize a specified loss function, typically using techniques like backpropagation through time$ (BPTT)$ or variants of gradient descent. The model learns to update the parameters of the $LSTM$ cells to make accurate predictions based on the patterns and dependencies in the training data. The $LSTM$ model's ability to retain and selectively update information over long sequences makes it a powerful tool for modeling and predicting sequential data. Its architecture has been widely used and has demonstrated state-of-the-art performance in various domains. 

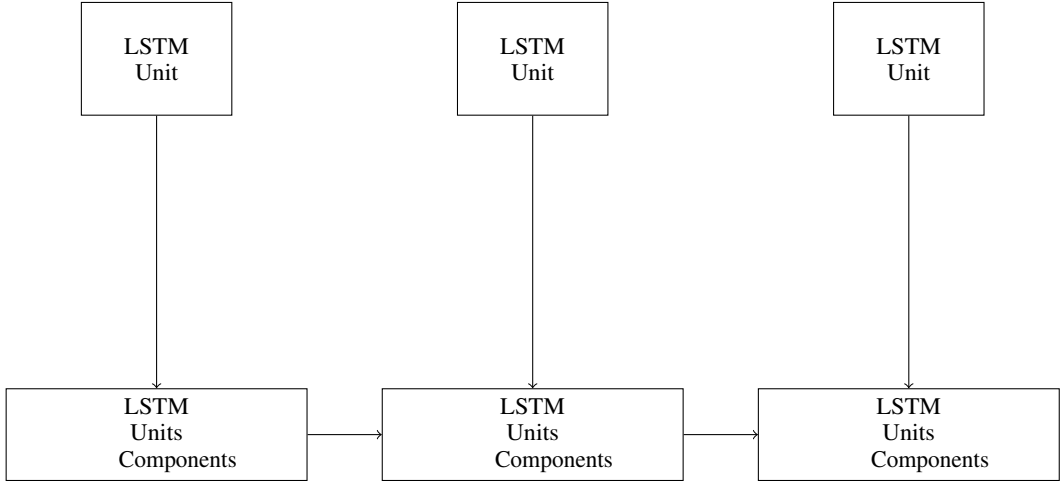
\begin{figure}[H]
\centering
\begin{tikzpicture}[node distance=5cm, every node/.style={draw,align=center, text width=1cm, font=\small}]
    \node (lstm1) [draw, minimum width=2cm, minimum height=1.5cm] {LSTM Unit};
    \node (lstm2) [draw, minimum width=2cm, minimum height=1.5cm, right of=lstm1] {LSTM Unit};
    \node (lstm3) [draw, minimum width=2cm, minimum height=1.5cm, right of=lstm2] {LSTM Unit};

    \node (comp1) [draw, minimum width=4cm, minimum height=1cm, below of=lstm1] {LSTM Units\\Components};
    \node (comp2) [draw, minimum width=4cm, minimum height=1cm, below of=lstm2] {LSTM Units\\Components};
    \node (comp3) [draw, minimum width=4cm, minimum height=1cm, below of=lstm3] {LSTM Units\\Components};

    \draw[->] (lstm1) -- (comp1);
    \draw[->] (lstm2) -- (comp2);
    \draw[->] (lstm3) -- (comp3);
    \draw[->] (comp1) -- (comp2);
    \draw[->] (comp2) -- (comp3);
\end{tikzpicture}
\caption{LSTM Model Architecture}
\label{fig:lstm_modell}
\end{figure}

In this part of the diagram in figure \ref{fig:lstm_modell}, the $LSTM$ unit is depicted as the core unit of the network. It consists of several key components, including the input gate, forget gate, output gate, and cell state. At each time step, these gates and the cell state are updated based on the input, the previous hidden state, and the previous cell state.

\subsection{Hermite Long-Short Term Memory}

The Hermite activation function is a non-linear activation function that can be used in $LSTM$ (Long Short-Term Memory) networks.  This algorithm consists of multiple layers, and each layer contains several $LSTM$ cells. Each of these cells has a cell state $(c_t)$ and a hidden state $(h_t$), which are calculated at each time step $t$. In this algorithm, Hermite activation functions are used.  The use of Hermite functions for solving differential equations in decision-making can also be found in other forms\cite{bib21}.

The Hermite activation function for an $LSTM$ layer can be written as: 

   \begin{equation}
  i_t = H(W_i \odot [h_{t-1}, x_t] + b_i),
  \end{equation}
   \begin{equation}
   f_t = H(W_f \odot [h_{t-1}, x_t] + b_f),
   \end{equation}
   \begin{equation}
   o_t = H(W_o \odot [h_{t-1}, x_t] + b_o),
    \end{equation}
   \begin{equation}
   g_t = H(W_g \odot [h_{t-1}, x_t] + b_g),
   \end{equation}
   \begin{equation}
    h_t = o_t \odot H(c_t),
    \end{equation}

where,

\begin{itemize}
    \item   $i_t$, $f_t $, and   $o_t$   represent the input gate, forget gate, and output gate, respectively.
    \item  $g_t$ is the gate activation.
    \item  $c_t $ is the cell state and $h_t $  is the hidden state.
    \item  $W_{i},W_{f},W_{o},W_{g}$ are the weight matrices.
    \item $b_{i},b_{f},b_{o},b_{g}$ are the bias vectors.
    \item $[h_{t-1},x_t]$denotes the concatenation of the previous hidden state and the current input.
    \item   $H$ is the Hermite activation function.
    \item ${H}_{n}(x)=\frac{1}{\sqrt{2^{n}n!}}e^{\frac{-x^{2}}{2}}(1-x^2),\quad n\geq0,\:x\in\mathbb{R}.$
\end{itemize}
Using these formulas and the Hermite function, we define an $LSTM$ neural network with the Hermite activation function. 
The equations provided in the question represent the Hermite activation function and the operation of the $LSTM$ cell. Here, input weights and biases are used to input the previous hidden state and the new input to calculate the output of each gate (input gate activation, forget gate activation, and output gate activation). Then, using the Hermite function and the input gate output, the new cell state $(c_t)$ is calculated based on the previous cell state $(c_{t-1})$ and the new input $(g_t)$. Finally, using the new cell state and the output gate output, the new hidden state $(h_t)$ is calculated.

In a typical $ LSTM$ network, the sigmoid function is used for the input and forget gates, which produce values between 0 and 1. However, this approach uses the Hermite activation function instead of the sigmoid function. The Hermite function is a non-linear function that has a "bell-shaped" curve and is specifically optimized for input range issues.

\begin{figure}[H]
\begin{tikzpicture}[node distance=2cm]

\tikzstyle{lstm} = [rectangle, draw, fill=blue!20, text width=3cm, text centered, minimum height=1cm]
\tikzstyle{arrow} = [thick,->,>=stealth]

\node[lstm] (cell1) {HLSTM Cell 1};
\node[lstm, below of=cell1] (cell2) {HLSTM Cell 2};
\node[below of=cell2, yshift=-1cm] (dots) {$\vdots$};
\node[lstm, below of=dots, yshift=-1cm] (celln) {HLSTM Cell N};
\node[lstm, left of=cell2, xshift=-3cm] (celln1) {HLSTM Cell N-1};
\node[lstm, right of=cell2, xshift=3cm] (celln2) {HLSTM Cell N+1};

\draw[arrow] (cell1) -- (cell2);
\draw[arrow] (cell2) -- (dots);
\draw[arrow] (celln1) -- (cell2);
\draw[arrow] (celln2) -- (celln2);
\draw[arrow] (dots) -- (celln);
\draw[arrow] (celln.west) -- ++(-1.5cm,0) |- (cell2.west);
\draw[arrow] (cell2.east) -- ++(1.5cm,0) |- (celln.east);
\draw[arrow] (cell1.west) -- ++(-1.5cm,0) |- (celln.west);
\draw[arrow] (celln.east) -- ++(1.5cm,0) |- (cell1.east);


\end{tikzpicture}
 \caption{Hermite $LSTM$ Model Architecture for this problem} \label{HermiteLSTM}
\end{figure}
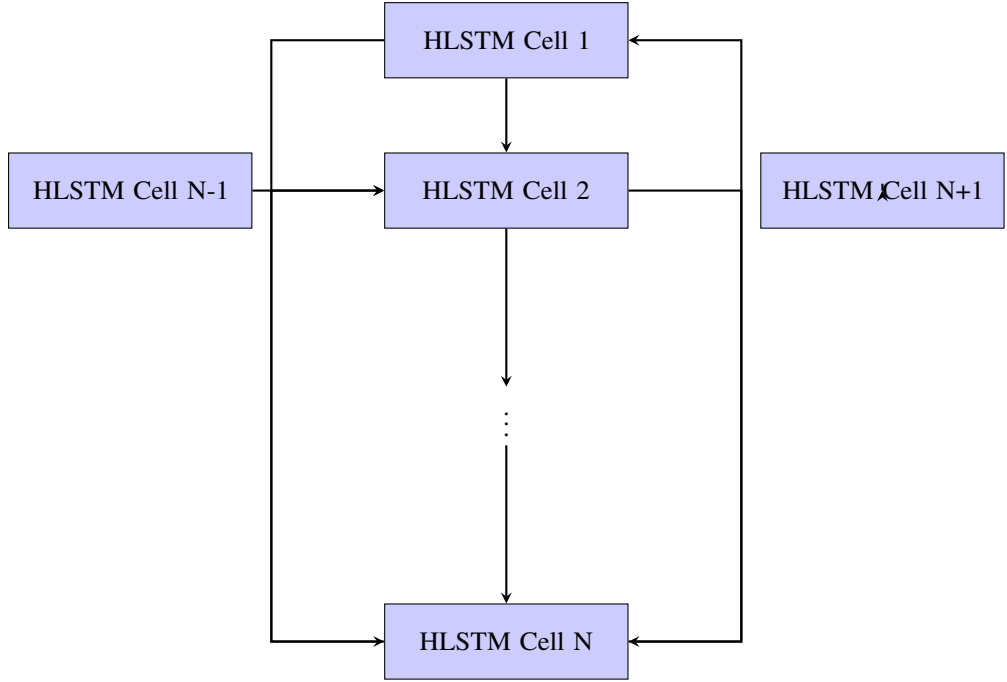

By using this approach, Hermite activation functions can be used instead of sigmoid functions in $LSTM$ neural networks, potentially leading to improved network performance and prediction accuracy. The architecture for this method for this problem is shown in figure \ref{HermiteLSTM}.

In this method, the general form of the time differential equations for an $LSTM$ model is as follows: 
\begin{equation}
    \frac{d\mathbf{h}_t}{dt} = \mathbf{f}(\mathbf{h}_t) \odot \mathbf{i}(\mathbf{h}_t) - \mathbf{o}(\mathbf{h}_t) \odot \mathbf{c}(\mathbf{h}_t),
\end{equation}

\begin{equation}
    \frac{d\mathbf{c}_t}{dt} = \mathbf{g}(\mathbf{h}_t) \odot \mathbf{i}(\mathbf{h}_t),
\end{equation}

Here, $\mathbf{h}_t$ represents the $LSTM$ state vector at a time $t$, and $\mathbf{c}_t$ represents the $LSTM$ memory vector at time $t$. The functions $\mathbf{f}$, $\mathbf{i}$, $\mathbf{o}$, are forget, input, and output cells respectively and $\mathbf{g}$ are activation functions that depend on $\mathbf{h}_t$ and $\mathbf{c}_t$ is the $LSTM$ memory vector at time $t$.

The input gate and forget gate in an $LSTM$ model play a crucial role in determining how the memory cell state is updated over time. Let's explore how these gates affect the memory cell state.

Input Gate:

The input gate regulates how much new information should be stored in the memory cell. It takes into account the current input at the given time step $(x(t)) $and the output of the previous time step$ (h(t-1))$. The input gate activation is computed using the Hermite functions:
\begin{equation}
    i(t) = H(W_i * x(t) + U_i * h(t-1) + b_i),
\end{equation}

here, $W_i$ and $U_i $are weight matrices, $b_i $is the bias vector, and $H$ represents the Hermite activation function.

The input gate activation $i(t)$ is mapping a value between 0 and 1, where 0 means "completely forget" and 1 means "completely remember". It determines how much new information will be added to the memory cell state.

Forget Gate:

The forget gate determines how much of the previous memory cell state should be retained or discarded. It considers the current input at the given time step $(x(t))$ and the output of the previous time step $h(t-1)$. The forget gate activation is computed using the Hermite function:

\begin{equation}
    f(t) = H(W_f * x(t) + U_f * h(t-1) + b_f),
\end{equation}

 $W_f$, $U_f$, and $b_f$ are weight matrices and a bias vector, respectively.

The forget gate activation $f(t)$ is also mapping a value between 0 and 1. A value of 0 means "completely discard" the previous memory cell state, while a value of 1 means "completely retain" it.

Cell State:

Combining the input and forget gates, we update the memory cell state $c(t)$ as follows:

\begin{equation}
    c(t) = f(t) * c(t-1) + i(t) * H(W_c * x(t) + U_c * h(t-1) + b_c)
\end{equation}

$W_c$, $U_c$, and $b_c$ are weight matrices and a bias vector, respectively. The term $H(W_c * x(t) + U_c * h(t-1) + b_c) $ represents the new information that could be stored in the memory cell.

The forget gate $f(t)$ scales the previous memory cell state $c(t-1)$, while the input gate $i(t)$ scales the new information. The gates work together to determine how much of the previous state to forget and how much of the new information to remember, allowing the $LSTM$ model to selectively update the memory cell state.

This mechanism enables the $LSTM$ model to retain important information over long sequences while allowing irrelevant or outdated information to be forgotten. It helps address the vanishing gradient problem and allows the model to capture long-term dependencies in the data.

\section{Results}\label{sec2}
In this section, we present the results obtained by solving the given system of differential equations \ref{eq3}, \ref{eq4}, \ref{eq5}, and \ref{eq6} using the proposed method. In this method, we used the Hermite functions as activation functions. The time interval considered in this method is from 0 to 10, with a total of 2025 iterations. We train this model by the Adam optimization algorithm, which is a popular choice for optimizing neural networks. Fig shows the comparison between the actual values and the predicted values. 
The neural network architecture of the Long Short-Term Memory $(LSTM)$ model is defined as follows:

\textbf{Architecture Parameters of the$ LSTM$ Model: }
\begin{itemize}
    \item  Number of $LSTM$ Units: The network consists of a single $LSTM$ layer with 71 units.
    \item Network Input: The network input is defined to be of dimensions (n\_samples, 1).
    \item Activation Function: In this model, converted Hermite functions are used as the activation function for the $LSTM$ and Dense layers.
\end{itemize}
\textbf{Model Layers:}
\begin{itemize}
    \item $ LSTM $Layer: A single $LSTM$ layer with 71 units is defined, processing inputs sequentially.
    \item Dense Layers: Two Dense layers with 20 and 26 neurons, respectively, are included for enhanced information processing and prediction of final outputs.
    \item Output Dense Layer: An output-dense layer with 3 neurons transforms the information into the final model output.
\end{itemize}
\textbf{Compilation and Model Training: }
\begin{itemize}
    \item  Loss Function: Mean Squared Error ($MSE)$ is used as the loss metric for error calculation.
    \item  Optimization Algorithm: The Adam optimization algorithm is employed for optimizing the network parameters.
\end{itemize}
\textbf{Training Epochs:} 
\begin{itemize}
    \item The model is trained for 2505 epochs in this implementation.
\end{itemize}
The $LSTM $architecture in this method is employed for modeling and predicting changes in view counts and user behaviors using a defined mathematical model. 

\begin{figure}[h]
\begin{centering}
\includegraphics[width=0.9\textwidth]{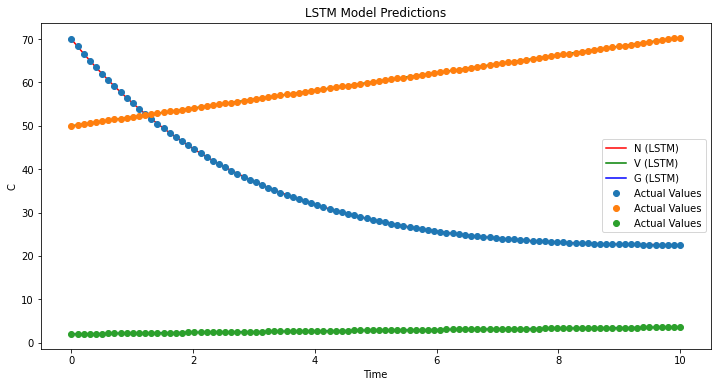}
\par\end{centering}
\caption{LSTM neural network}\label{actualpredicate}
\end{figure}

The plot showcases the changes in these values over time. In the figure \ref{actualpredicate} is shown as prediction vs actual and the blue lines represent the model's predictions for $N$, $V$, and$ G$ values, respectively. The points depicting the numerical solutions of the differential equation represent the actual values of $N$,$ V$, and $G$ over time. This plot gives us insights into how well our model has captured the underlying patterns and trends in the data.

\begin{figure}[h]
\begin{centering}
\includegraphics[width=0.9\textwidth]{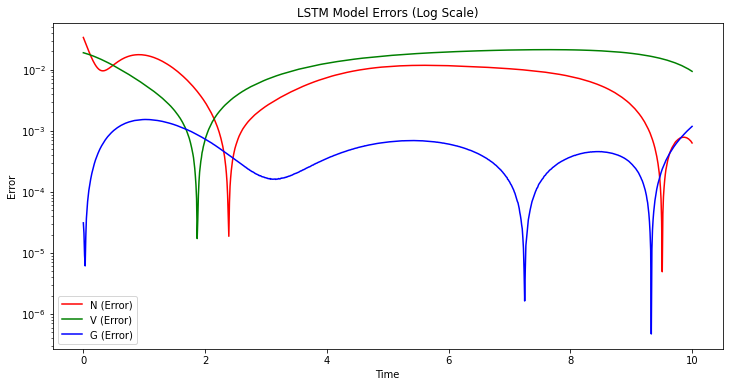}
\par\end{centering}
\caption{LSTM neural network}\label{error}
\end{figure}

We demonstrated the distribution of errors in our predictions. The error distribution plot provides information about the spread and concentration of errors across the dataset in the figure \ref{error}. Analyzing this distribution can help us identify specific patterns or outliers where the model might struggle to make accurate predictions. By employing the Adam optimization algorithm and presenting these visualizations, we aim to showcase the effectiveness and performance of our proposed method in solving the given system of differential equations.

\section{Conclusion}\label{sec13}

The solution of the specified differential equation in the question enables an examination of the changes in the number of views on a character over time. These equations \ref{eq3},\ref{eq4}, \label{eq5}, and \ref{eq6} is a mathematical model that models the number of views based on the fixation rate and separation rate.

In the differential equation \ref{eq3}, $\frac{dN(t)}{dt}$ represents the changes in the ratio of views over time. These changes are equal to the difference between the fixation rate $(\lambda G(t))$ and the separation rate $(\mu N(t))$. We showed that if the fixation rate is greater than the separation rate, the number of views increases, and if the separation rate is greater than the fixation rate, the number of views decreases.

To solve this differential equation, we needed initial conditions. The initial conditions included the number of views at the initial time (e.g., $t=0$), denoted by. We needed to know the fixation rate and the separation rate.

With the initial conditions and information about the fixation and separation rates, we solved the differential equation and obtained the results. These results help us understand and improve user interface design, games, or related systems.

In the differential equations \ref{eq3},\ref{eq4}, \ref{eq5}, and \ref{eq6}, other factors such as the character's speed relative to the player and the character's quality (good or bad) $(G(t))$ are considered. The character speed and quality changes are also modeled using the corresponding differential equations.

By solving these differential equations, we examined the impact of various factors on the number of views and user behavior.  We analyzed the interaction between these factors and obtained insights useful for improving design and user experience. We gained the number of views and user behavior that can help us understand and enhance systems, interfaces, and user experience.
In summary, we can state the following results

 \begin{enumerate}
     \item \textbf{Development and Implementation of the Model:} we explain the $LSTM$ network architecture and the activation function (Hermite functions) used in the present model, and discuss important model parameters such as the number of layers, the number of neurons, the cost function, and the optimization algorithm.

     \item \textbf{Preparation of Training Data:} we describe how the training data is generated from the differential\_equations function and how it is used as input for the model. We also provide information about the parameters of the differential equations, such as $\lambda$, $\mu$, $k$, and $m$.

     \item \textbf{Model Training and Evaluation:} we explain how the model is trained using the compile and fit functions, and display the accuracy (loss) graph during the training process.

     \item \textbf{Prediction of Future Values:} we demonstrate the results of predicting future values using the predict function. The results compare with actual data and explain how well the model could provide better predictions than the actual data.

 \end{enumerate}



\end{document}